# HISTORICAL CHINESE EFFORTS TO DETERMINE LONGITUDE AT SEA

**Richard de Grijs**
*Department of Physics and Astronomy, Macquarie University,
Balaclava Road, Sydney, NSW 2109, Australia*
Email: richard.de-grijs@mq.edu.au

**Abstract:** High-level Chinese cartographic developments predate European innovations by several centuries. Whereas European cartographic progress—and in particular the search for a practical solution to the perennial 'longitude problem' at sea—was driven by persistent economic motivations, Chinese mapmaking efforts responded predominantly to administrative, cadastral and topographic needs. Nevertheless, contemporary Chinese scholars and navigators, to some extent aided by experienced Arab navigators and astronomers, developed independent means of longitude determination both on land and at sea, using a combination of astronomical observations and timekeeping devices that continued to operate adequately on pitching and rolling ships. Despite confusing and speculative accounts in the current literature and sometimes overt nationalistic rhetoric, Chinese technical capabilities applied to longitude determination at sea, while different in design from European advances owing to cultural and societal circumstances, were at least on a par with those of their European counterparts.

## 1 PREAMBLE

Prior to the development of sufficiently accurate timing devices, determination of one's position at sea posed a significant practical problem for transoceanic voyages. While latitude determination only requires one to measure the height of the Sun above the horizon at its local meridian passage (i.e., at local noon), or that of Polaris, the North Star, at night, accurate longitude determination relies on knowing one's local time precisely with respect to that at a reference location. Most accounts of attempts to find practically viable solutions to the thorny 'longitude problem' tend to focus on European developments in the fifteenth to eighteenth centuries (for a recent book-length review, see de Grijs, 2017). Early progress made elsewhere often does not attract more than cursory commentary, despite significant advances in pre-Columbian oceanic exploration achieved by, e.g., Polynesian, Arab and Chinese navigators.

Unfortunately, the scholarship and discourse pertaining to early Chinese[1] cartography and the country's historic transoceanic voyages is rife with speculation (e.g., Finlay, 2004; Thompson, 2008; Robinson, 2010) and sometimes overtly nationalistic rhetoric (e.g., Sen, 2016; Benabdallah, 2021; and references therein). In this paper, I review early Chinese cartographic achievements in the context of efforts to determine one's longitude at sea. I focus almost entirely on actual quotations from historical sources. My aim here is to provide a 'clean' history of Chinese efforts to solve the longitude problem. My secondary aim is to compare Chinese developments with European approaches to longitude determination, specifically to showcase the high level of Chinese capabilities at much earlier times.

Early Chinese successes in cartography culminated in the efforts of the Imperial official Péi Xiù[2] (裴秀; 223–271 CE: see Figure 1[3]), who is often considered the founder of traditional Chinese cartography. He is credited with defining the principles of scale, grid and triangulation, and also with standardising the representation of uneven heights on a plane surface. Yet, the ancient Chinese did not develop world maps. The earliest-known Chinese 'world map' is the *Map of China and the Barbarian Countries* (1136). However, this map is in reality a national map that only shows those areas of the Korean peninsula and Vietnam immediately adjacent to Chinese territory, with the names of other countries indicated in their approximate locations (e.g., Song and Chen, 1996). In fact, ancient Chinese maps were often not based on survey measurements but rather on a critical assessment of earlier maps, literary sources and personal observations (Yee, 1994; Sun, 2009).

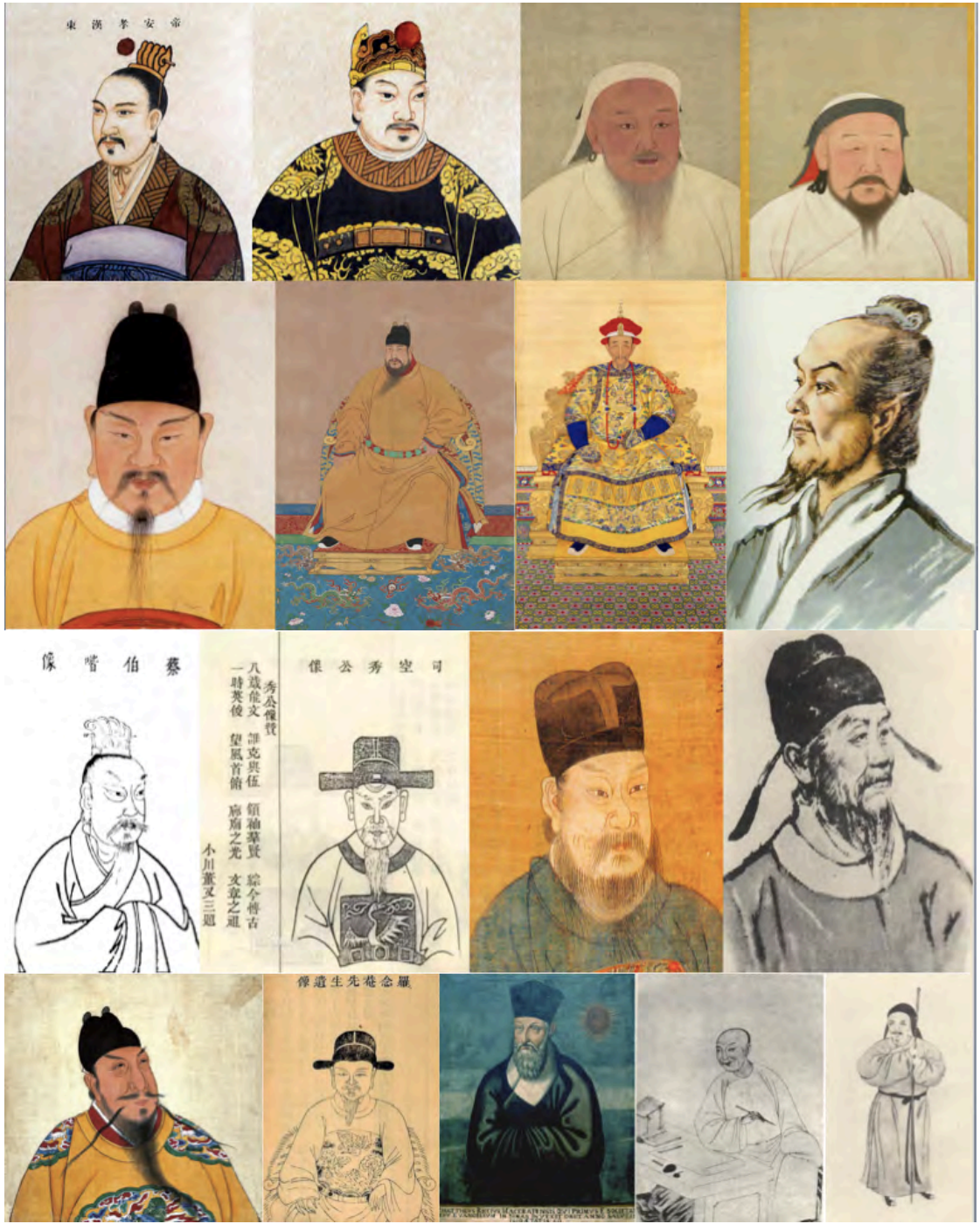

**Figure 1.** Portraits of the main characters driving the innovations described in this paper, ordered from left to right and from top to bottom by date or birth. The top two rows represent emperors (and a Korean king); the bottom two rows are scholars, navigators and officials of some importance to the story. Individuals depicted include *(i)* Emperor Ān (94–125 CE), *(ii)* Emperor Shùn (115–144 CE), *(iii)* Ghengis Khan (1162–1227), *(iv)* Kublai Khan (1215–1294), *(v)* Yǒnglè Emperor (1360–1424), *(vi)* Hóngxī Emperor (1378–1425), *(vii)* Kāngxī Emperor (1654–1722), *(viii)* Zhāng Héng (78–139 CE), *(ix)* Cài Yōng (132–192 CE), *(x)* Péi Xiù (223–271 CE), *(xi)* Yélǜ Chǔcái (1190–1244), *(xii)* Guō Shǒujìng (1231–1316), *(xiii)* Zhèng Hé (1371–1433/5), *(xiv)* Luó Hóngxiān (1504–1564), *(xv)* Matteo Ricci (1552–1610), *(xvi)* Hú Wèi (1633–1714) and *(xvii)* Lǐ Zhàoluò (1769–1841). *Figure credits*: Wikimedia Commons, except for *(i)* Sogou; *(ii)* and *(x)* KK News; *(xi)* Historium; *(xii)* Kekenet; *(xiii)* National Museum of China. All images are in the public domain, except for *(viii)*: CC-BY-SA.

The Taoist monk Zhū Sīběn (朱思本; 1273–1337), who is best known for having made the first maritime voyage from China to the Atlantic Ocean, applied a grid pattern to his *Terrestrial map* (*Yútú*), to facilitate plotting directions and distances. Zhū omitted some foreign countries from his map owing to a lack of reliable information:

> As for the various foreign lands southeast of the South Sea and northwest of the great desert [Mongolia], although they at times send tribute to the court, their distance precludes investigation. Those who speak of them are unable to be specific. Those who are specific cannot be trusted. (Zhū Sīběn; in Luó, 1579: 1a, note 103; own translation[4]).

In the preface to his *Enlarged Terrestrial Atlas* (ca. 1555: see Figure 2), Luó Hóngxiān (羅洪先; 1504–1564), Zhū's Míng-era successor, explicitly attributed the accuracy of the *Yútú* map to the use of a grid pattern:

> His maps employ the method of drawing squares for measuring distance, and as a result, their depiction of reality is reliable. Thus, whether one divides or combines them, East and West match each other without incurring any discrepancy. (Luó, 1579: preface, 2a; own translation).

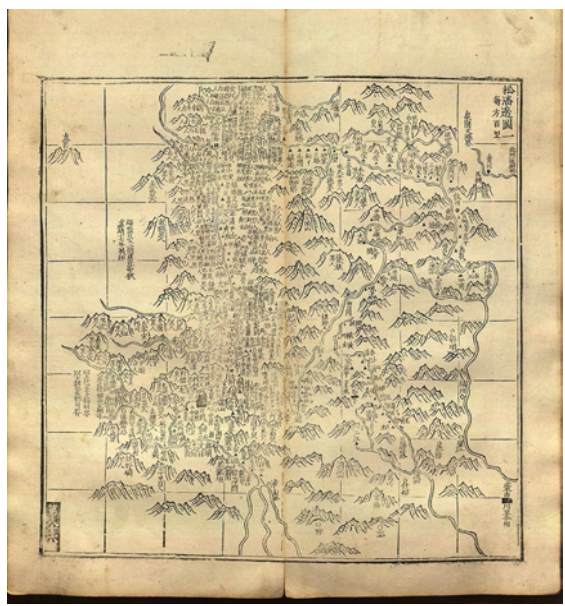

**Figure 2.** Excerpt from Luó Hóngxiān's *Enlarged Terrestrial Atlas* (ca. 1555), which includes a Chinese grid. The side of each grid cell represents 100 miles. (Courtesy: Library of Congress, Geography and Map Division, digital ID lcnclscd.2008623187; public domain.)

**2 EARLY CHINESE CARTOGRAPHY**

Early Chinese maps were predominantly meant for domestic and local administrative purposes, not for exploration. It seems that the ancient Chinese did not entertain any ideas related to the existence of distant lands across the vast Pacific Ocean. In fact, although evidence exists of Chinese monks having travelled to Japan more than a thousand years ago (e.g., Wei, 2005), that country does not appear on any early Chinese maps (Song and Chen, 1996). Similarly, Admiral Zhèng Hé (鄭和; 1371–1433/5) never ventured eastwards into the Pacific during any of the seven voyages he undertook during the Míng dynasty (1368–1644) (see Section 7; see also Duyvendak, 1939; Mills, 1970; Dreyer, 2007; but see Ptak, 2007, for scathing criticism of the latter publication).

(For period reference, Table 1 provides a timeline of the ruling Chinese dynasties over the course of the historical developments discussed in this paper.)

The foundation of Chinese cartography was, instead, based on the country's agricultural needs. If anything, administrative policies encouraged agricultural enhancements while restricting commerce and discouraging overseas trade. Maps were developed primarily for land distribution, management of tax collection, military campaigns, urban defence, river and terrestrial transport, and water conservation. As a result, early Chinese maps were highly accurate, large-scale maps of small areas that were meant for cadastral, administrative and topographic purposes (e.g., Chao, 1978). As such, there was no need to determine accurate latitudes and longitudes, nor to correct map projections for the effects of the Earth's curvature (Song and Chen, 1996). Although they represented distances and topographical features very accurately, such early administrative or military maps often did not include grid patterns (e.g., Zhang, 2015: Chapter 2, note 48).

**Table 1**: Chinese dynastic history

| Dynasty | Period | Notes |
|---|---|---|
| Hàn | 206 BCE – 220 CE | Later (Eastern) Hàn, 25–220 CE |
| Wèi | 220–265 CE | |
| Jìn | 265–420 CE | |
| Northern and Southern Dynasties | 386–589 CE | (*not covered in this paper*) |
| Suí | 581–618 CE | |
| Táng | 618–690 CE, 705–907 CE | |
| Five Dynasties Period, Ten Kingdoms | 907–960 CE | (*not covered in this paper*) |
| Sòng | 960–1271/9 | |
| Yuán | 1271/9–1368 | |
| Míng | 1368–1644 | |
| Qīng | 1644–1912 | |

Nevertheless, Chinese navigators appear to have developed a rudimentary means of longitude determination during the Yuán (1271/9–1368) and early Míng dynasties (see Section 7). As we will see, ancient Chinese cartographers were already well versed in mapping their territory, which may explain why the Venetian merchant and explorer Marco Polo (1254–1324) was familiar with the longitude concept upon his return from China in the thirteenth century (e.g., Olshin, 2014).

It is often claimed that medieval Islamic developments influenced the development of Chinese geography under the thirteenth-century Mongol Empire and its successor state, the Yuán dynasty. In fact, although the earliest influence of Islamic astronomy in China dates from 961 CE (J.-J. Chen, 1996: 52–66; Shi, 2014), i.e., during the Sòng dynasty (960–1271/9), it took until the rise of the Mongol Empire before Islamic influences in North Asia extended to cartography. However, there is an increasing body of evidence suggesting that home-grown Chinese cartographic advances predate those influences by more than a millennium.

Indeed, the use of maps covered by rectangular (or square) grid patterns probably originated in China (Needham, 1986: 106–107; for an opposing view, see Yee, 1994), most likely with the work of Zhāng Héng[5] (張衡; 78–139 CE) (Nelson, 1974: 357–362; de Crespigny, 2007: 1049–1051; for a biography, see Trimble et al., 2007: 1264–1265). Zhāng, the leading polymath of the Later (Eastern) Hàn dynasty (25–220 CE), was an engineer, meteorologist, geologist and philosopher based in the Hàn capital of Cháng'ān (present-day Xī'ān; the eastern extremity of the Maritime Silk Road). He was instrumental in advancing fields as disparate as astronomy, cartography, mathematics, philosophy and literature.

**3 ZHĀNG HÉNG'S PIONEERING MAPS**

Most relevant for our exploration of early Chinese cartography, Zhāng designed a water-driven armillary sphere (see Figure 3), as well as a celestial globe, to track the movements of stars and planets, in support of his efforts to develop the *hùntiān* theory of cosmology. He published two leading treatises on that subject, *The Spiritual (Sublime) Constitution of the Universe* (120 CE) and *Commentary (Treatise, Notes) on the Armillary Sphere* (ca. 117 CE). In the latter manuscript, Zhāng described the geometry of the universe, comparing the heaven with the (spherical) shell of a hen's egg with Earth at its centre (e.g., Trimble et al., 2007).

Given the assumed sphericity of the heaven, spherical coordinates could be defined. Chinese celestial (equatorial) coordinates consisted of right ascension and (north) polar distance, expressed in *dù*, the angular distance covered by the Sun's projected daily motion. To define right ascension, the hour circles passing through the principal stars of the 28 traditional astrological 'lunar mansions' (lodges)—segments of the ecliptic traversed by the projected lunar orbit—were used.

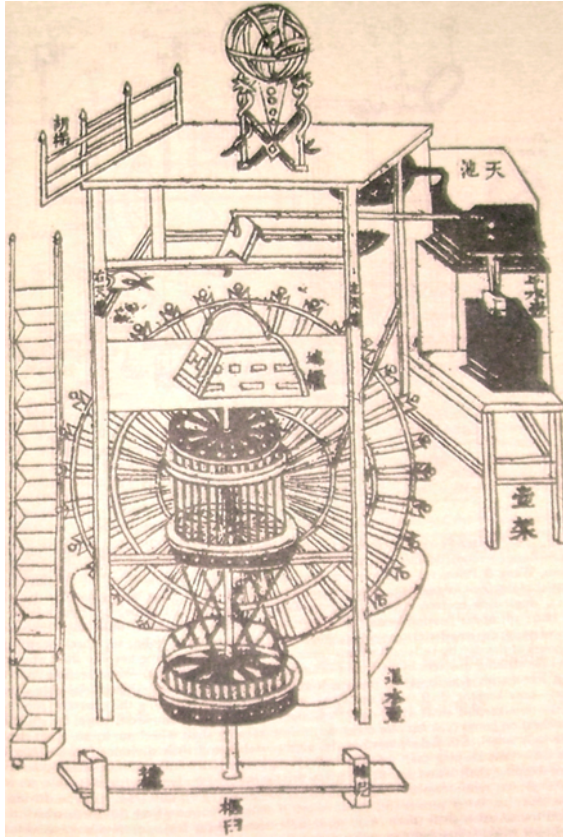

**Figure 3.** Original diagram of Sū Sòng's (1020–1101) astronomical clock tower, featuring (top) an armillary sphere powered by a waterwheel according to Zhāng Héng's design. (Wikimedia Commons; public domain.)

Zhāng also developed the idea of lunar eclipses as geographic aids and invented quantitative cartography, allowing him to apply longitude and latitude grids to maps. He was appointed as an Imperial official by Emperor Ān of Hàn (漢安帝; reigned 106–125 CE). He was subsequently promoted to Director of the Imperial Bureau of Astronomy and Calendrics, a role to which he was reappointed by Emperor Ān's successor (not including the very short-lived reign of Emperor Shǎo; 少帝), Emperor Shùn of Hàn (漢順帝; reigned 126–144 CE).

Zhāng's maps have been lost, although his *Dìxíng tú* topographical map is recorded (Zhāng, 847 CE) as having been used until at least the Táng dynasty (618–690 and 705–907; see Yee, 1994). However, we can obtain some interesting insights from descriptions in Zhāng's biography by the scholar Cài Yōng (蔡邕; 132–192 CE), who stated that Zhāng Héng "… cast a network about heaven and earth and reckoned on the basis of it." (Temple, 1989: 30).

Zhāng's pioneering work, collected in his treatise *Discourse on New Calculations* (ca. 116 CE), greatly influenced that of his Wèi (220–265 CE) and Jìn dynasty (265–420 CE) successor, the cartographer Péi Xiù whom we encountered already. Péi was the first scholar to include a mathematical grid reference and a graduated scale on his maps for improved accuracy in distance estimates (Needham, 1989: 538–540; Hsu, 1993). He adopted the grid as one of his 'six principles of scientific cartography'—comprising scale, bearing, distance, square/oblique, high/low, curved/straight—and 'division by *lǐ*' (a *lǐ* is a little shorter than 600 m, although it is often equated to 500 m). These principles had been developed in more rudimentary fashion by his predecessors, such that …

> When the principle of the rectangular grid is properly applied, then the straight and the curved, the near and the far, can conceal nothing of their form from us. (Harvey, 1980: 133–134).

However, Yee (1994) argues that Péi's square grid pattern was practically different from the pattern included on later Chinese maps, including the famous Sòng dynasty stone-engraved grid map from 1137, the *Map of the Tracks of Yǔ*: see Figure 4. This latter map is covered by 5110 squares, with each grid cell representing approximately 100 *lǐ* on a side, resulting in an overall scale of approximately 1:5,000,000 (e.g., Zhang, 2015: Chapter 2). The *Map of the Tracks of Yǔ* represents the country's river systems, lakes and coastlines at unprecedented precision for the times. Yee (1994) suggests that Péi's map instead used a single reference point, which would facilitate and regulate 'backsighting' as a means of checking one's bearings:

> Regulated sighting is a means of rectifying the configuration of this place and that [that is, relative position] ... If a map has proportional measure [scale] but lacks regulated sighting, then, though it may be correct in one corner, it will fail in other places. (Péi, 648 CE; translation: Yee, 1994).

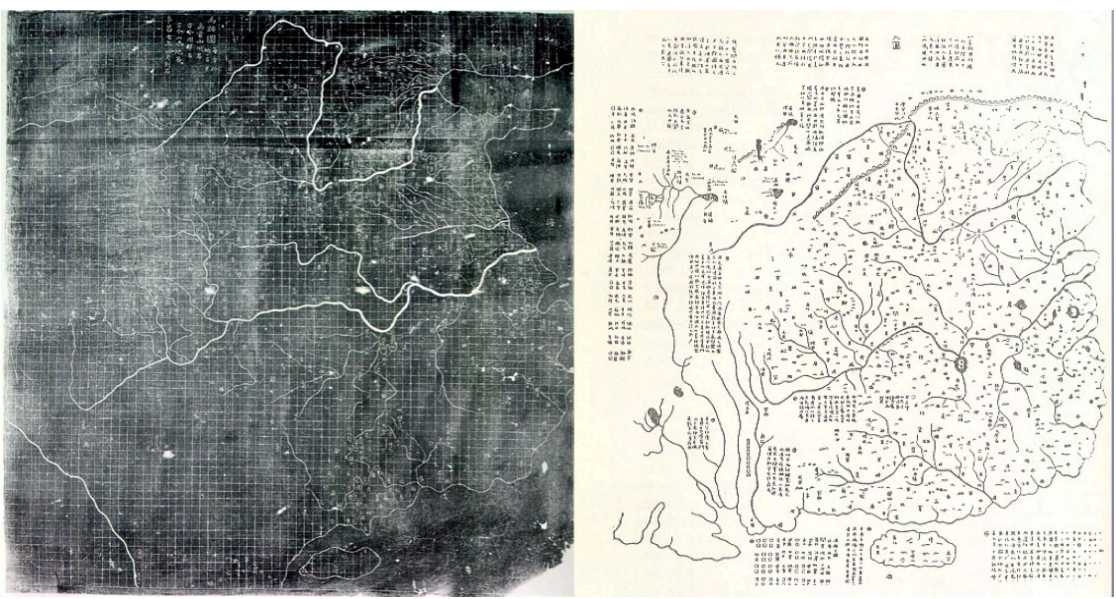

**Figure 4.** *(left)* Stele-engraved *Map of the Tracks of Yǔ* (1137) (Needham, 1959: Plate LXXXI; via Wikimedia Commons, out of copyright). *(right) Map of China and the Barbarian Countries* (Needham, 1959; via Wikimedia Commons, out of copyright).

In any case, the introduction of a grid pattern and establishing the relative positions of places of interest on land is not the same as determining one's longitude on the open seas. In fact, ancient Chinese maps—including he *Map of the Tracks of Yǔ*—may have featured square grid patterns, but the grid was not meant to operate in the same sense as the mathematical *graticule* pattern on modern (Western) maps. Instead, Chinese grids appear to have been superimposed on maps almost arbitrarily. Grids were primarily meant to help map readers determine distances, directions and land areas (for a detailed discussion on the origin of the grid pattern, see Yee, 1994).

**4 LUNAR ECLIPSE OBSERVATIONS**

Ancient Chinese cartographers were well versed in determining their positions accurately on land, using triangulation and observations of the night sky, but longitude determination at sea required further developments. By the Táng dynasty, Chinese astronomers had mastered the skills to determine their latitude, while the world's earliest meridian measurement was also achieved around this time (e.g., Zhu, 2001).

Longitude determinations using lunar eclipses were most likely introduced to China by Arab astronomers during the Mongol Empire and the Yuán dynasty (e.g., van Dalen, 2002; Zhang, 2002). However, systematic observations of lunar eclipses and attempts at predicting them had started already early in the third century CE. Eclipse timings were usually obtained with the aid of clepsydras (water clocks). Chinese techniques spread across East Asia, particularly to Japan and the Korean peninsula. High-quality eclipse records from across East Asia became common after about the year 1000, reaching a highest accuracy on the order of one-quarter of an hour by the end of the thirteenth century (Song, 1370).

During his 1220–1221 campaigns in Tashkent and Samarkand (in present-day Uzbekistan), the Mongol ruler Ghengis Khan (1162–1227; reigned 1206–1227) requested that his Khitan consultant, the high official 'Longbeard' Yélǜ Chǔcái (耶律楚材; 1190–1244), an expert in astronomy and astrology, engage with local Muslim astronomers so as to obtain predictions of two lunar eclipses. Yélǜ's own eclipse timing predictions turned out to be more accurate than those of his local colleagues (Yabuuti and van Dalen, 1997; Shi, 2014).

Nevertheless, these latter scholars taught him the concept of time differences between different geographic longitudes. Yélǜ incorporated this idea into his own work, the *Gēngwǔ Epoch System of Calendrical Astronomy for the Western Expedition*, which is recorded[6] in the *History of Yuán* (Song, 1370). This *Western Expedition Calendar with Epoch Year Gēngwǔ* (that is, 1210) is almost identical to the *Dàmíng lì*, the *Great Enlightenment Calendar* of the Jìn dynasty (Yabuuti and van Dalen, 1997), except for a correction for the use of Samarkand as geographic reference. Yélǜ explained in his section on the lunar position,

> Making *Xún sī gānchéng* [Samarkand] the standard, we put the distance in *lǐ*s on the counting board and multiply it by 4359. Moving the [decimal] position backwards, we divide by 10,000 and make *fēn*s [one ten-thousandth]. These are called the *lǐ difference*. (Yabuuti and van Dalen, 1997: 12).

Here, the *lǐ difference* is the geographic correction at Samarkand with respect to a reference location in mainland China.

Yuán dynasty Chinese astronomers were skilled in using lunar eclipses to determine their local longitude, although this was a rather cumbersome process (e.g., de Grijs, 2017: Chapter 2). It required synchronous time measurements across large distances. The lunar eclipse method applied by Chinese astronomers required a base observatory in China. For a given lunar eclipse, observers at both the base observatory and the location of interest identified bright stars that passed through the local meridian at the moment the Moon started to reappear after the eclipse. Upon the astronomers' return to the base observatory, both records were compared and a second measurement was obtained to determine the exact time difference of the meridian passages of both stars. The measured time difference corresponded directly to the difference in longitude between both locations.

Chinese astronomers had long known about the Moon's irregular motion, the 'inequality of lunation'. The orbital irregularities had first been discovered by the astronomer Jiǎ Kuí (賈逵; 30–101 CE) during the Eastern Hàn era (e.g., Sun, 2000). The Moon's apparent motion takes it 360 degrees around the Earth to its original position among the background stars in approximately 27.3 solar days. This motion corresponds to an average of 13 degrees per day, or slightly more than half a degree per hour. While the 'fixed' stars appear to move westwards because of the Earth's rotation, the Moon appears to describe a retrograde orbit with respect to the fixed stars, covering approximately half a degree per hour in the eastward direction. The Moon appears to speed up when going towards the Sun and it seems to slow down when retreating. The change in orbital motion also depends on the Earth's orbital position.

Having noticed the irregular nature of the lunar motion, Chinese astronomers used a variety of interpolation methods to predict the Moon's 'equation of time', leading to accurate predictions of the Moon's motion across the sky through the 'equal interval second difference method'—also known as the piecewise parabolic interpolation—proposed by Liú Zhuō[7] (刘焯; 544–610 CE) (Qu, 2000). This method was further improved upon during the Yuán dynasty, by Guōshǒujìng (郭守敬; 1231–1316; see Section 5), using third-degree interpolation—although for the *solar* motion, whose equation of time had been determined by the astronomer Zhāng Zǐxìn (張子信; ca. 560 CE) during the Suí dynasty (581–618 CE; Qu, 2000).

By simply measuring the angular distance on the sky between the Moon when it crosses the local meridian and a given star, one can calculate the local longitude. This requires only access to a sextant (or, in China, a 'star board'; see Section 7), without the need for an accurate clock. This thus implies that Chinese astronomers would have been able to accurately calculate lunar positions on the sky since the Yuán dynasty. In turn, this would have allowed them to compile ephemeris tables listing the positions on the sky of the Moon and a set of reference stars throughout the year. Armed with such tables and with access to a star board, they could have been able to determine their longitude at sea fairly accurately (cf. de Grijs, 2017: Chapter 2; Jin, 1996; see also Section 7). This method is, in fact, equivalent to the 'lunar distance method' employed by European astronomers several centuries later (e.g., de Grijs, 2020).

However, one might wonder how effective the use of eclipses would have been in the general context of longitude determination, given that these are rare events. During any calendar year, up to seven eclipses may occur, either four solar and three lunar or five solar and two lunar eclipses. Since lunar eclipses can be seen from the entire half of the Earth facing the Moon while solar eclipses are only visible from a narrow geographic path, at any given location on Earth one can see more lunar than solar eclipses during any time period. Still, their frequency is clearly insufficient to efficiently generate large-scale longitude catalogues.

## 5 GUŌ SHǑUJÌNG

We learn from the *Book of Astronomy* (1383) in the *History of Yuán* that the Arab astronomer ("… from the western regions …"; Zhu, 2001: 360) and the founder of the Islamic Astronomical Bureau, Jamal al-Din Muḥammad ibn Ṭāhir ibn Muḥammad al-Zaydī al-Bukhārī (ca. 1255–1291; van Dalen, 2007), introduced the *Theory of Earth Being Round* to the Mongol Emperor and his court.

In 1267, al-Din al-Bukhārī—better known by his Chinese name Zāmǎrǔd(d)īn or Zhāmǎlǔdīng (匝馬魯丁)—also built the first wooden terrestrial globe covered by a longitude and latitude grid and featuring a correct sea:land ratio of 70:30. He offered his globe to Guō Shǒujìng, the Chinese astronomer, engineer and mathematician whom the Jesuit astronomer Johann Adam Schall von Bell (1591–1666) called the 'Tycho Brahe of China'. Nevertheless, it appears that these foreign influences did not have a significant effect on traditional Chinese cartography (Zhu, 2001).

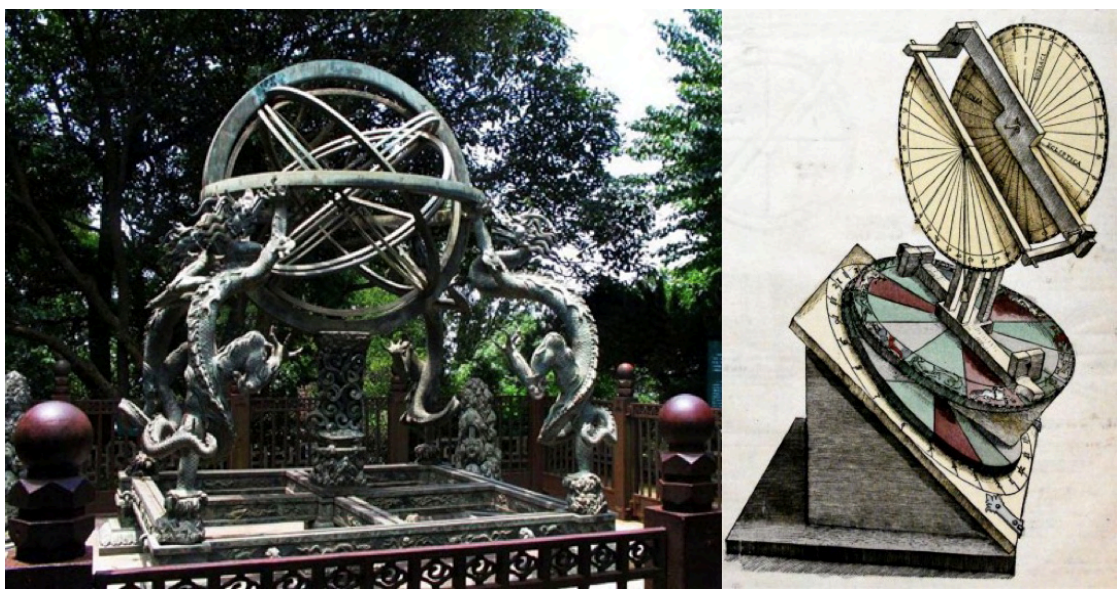

**Figure 5.** *(left)* Míng-era armillary sphere at Nánjīng Observatory (1437). It consists of three graduated ring systems, from which the equatorial, ecliptic and horizontal coordinates of celestial bodies could be measured. (Courtesy: Purple Mountain Observatory, Chinese Academy of Sciences). *(right)* Medieval torquetum designed to convert among coordinate systems. (Wikimedia Commons; public domain.)

As usual at the start of a new dynasty, in 1276 the first Yuán Emperor, Kublai Khan (1215–1294), assigned the compilation of a new and improved calendar to his court astronomers. Guō Shǒujìng was tasked with calendar reform, ultimately aiming to unite the Emperor's northern and southern territories. Among Guō's first notable achievements was the realisation that the armillary sphere at the Imperial Observatory in Nánjīng, the Yuán capital, had been constructed for a geographic latitude of 35°N, the latitude of the Northern Sòng capital, Kāifēng.

He proceeded to construct a dozen new instruments for the observatory, including a number of portable instruments for fieldwork and a new torquetum (turquet) which featured the positions of the main celestial bodies (the Sun, the Moon, the ecliptic, Polaris and other bright stars) for reference and which could be employed to determine longitudes based on the progressive daily difference between sidereal and either solar or lunar time.[8] Figure 5 shows (left) a Míng-era armillary sphere at Nánjīng Observatory and (right) an example of a medieval torquetum.

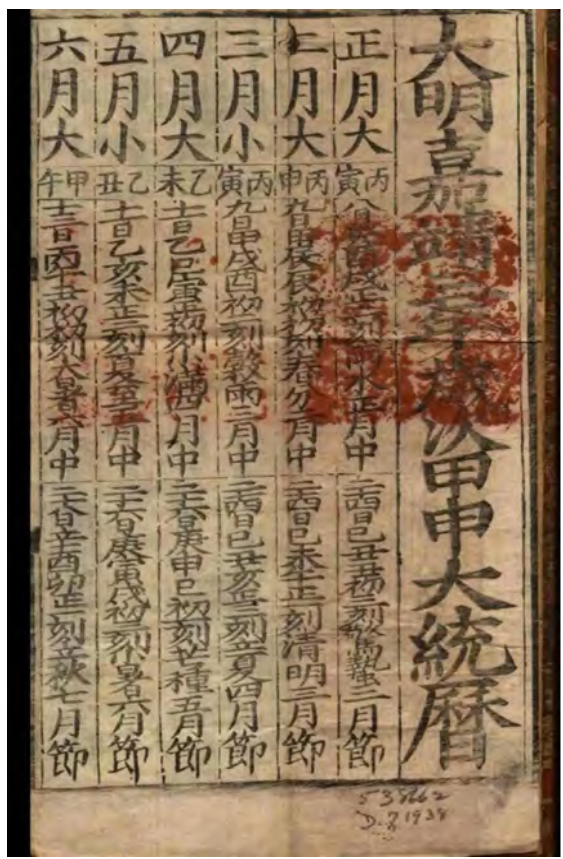

**Figure 6.** Excerpt from Guō Shǒujìng's *Shòushí* calendar. (Courtesy: baike.com; public domain.)

Guō had access to accurate water clocks. By 1276, Chinese engineers had developed compensating mechanisms for reduced water flow—and hence water pressure—as time progressed, for differences in air pressure, and for different water temperatures and salinity levels. Chinese water clocks were calibrated based on the number of drips between subsequent solar meridian passages. Guō and his colleagues completed the new Yuán calendar—*Shòushí(lì)* ('measuring time for the public': see Figure 6)—in 1280. They adopted the winter solstice of that year as the calendar's reference epoch.

Guō and his team had conducted numerous complicated calculations (e.g., to correct for variations in the speed of the projected solar motion) as well as astronomical observations across a wide area (latitudes, 15–65°N; longitudes, 102–128°E) to eventually determine the length of the (tropical) year at 365.2425 days—identical to the length of the year adopted for the Gregorian calendar in 1582—and that of the lunar month at 29.530593 days (Li and Zhang, 1996). The discrepancy between Guō's tropical year and the actual period of one rotation of the Earth around its axis was a mere 26 seconds.

Guō's *Shòushí* calendar contained astronomical data, including for 1461 bright stars, from thousands of observations, which had taken numerous astronomers many years to compile. Based on these data, comets, eclipses and the times of sunrise and sunset, as well as those of moonrise and moonset, could be predicted many years in advance. The *Shòushí* calendar also included the projected positions of the Sun, the Moon and the planets relative to the standard set of bright reference stars and to each other. The wealth of data allowed skilled mathematicians and astronomers to calculate the main celestial events for any day and almost any location. Specifically, the data were sufficient to calculate one's longitude[9] based on either the progressive daily difference between sidereal and solar time, lunar eclipses or the lunar distance method, that is, on the basis of the projected angular distances between pre-defined bright stars (or planets) and the lunar limb at a given time.

For navigation purposes, Guō's calendar established a common framework of longitude and latitude. The positions of all stars in the northern hemisphere could be fixed relative to Polaris, which in turn allowed the production of universal star maps for common use by the country's navigators. In addition, navigators were provided with extensive star tables, equivalent to what we now refer to as almanacs. Thus, all navigators were taught to sail on the basis of a common system of longitude and latitude, using a uniform set of stars (with agreed-upon names) to base their bearings on.

**6 EXPANDING CHINESE INFLUENCE**

Circumstantial evidence suggests that Chinese navigators may have mastered some degree of longitude determination already in the fourteenth century. As part of an expedition undertaken from 1334 to 1339, a traveller from Quánzhōu—Wāng Dàyuān (汪大淵; fl. 1311–1350)—successfully made a transoceanic voyage from Mozambique to Sri Lanka using a combination of compass and stellar position measurements called *Guò yáng qiān xīngshù* ('star orientation') to determine his ship's latitude and longitude (Wang, 2004a; Morton and Lewis, 2005: 128).

Meanwhile, the Mongol Emperors were keen to engage with Arab astronomers and visiting navigators alike (e.g., Zhang, 2002). The Yuán court ordered officials in Quánzhōu, the official starting point of the Maritime Silk Road during the Yuán dynasty, to collect Arab compass navigation charts from visiting navigators. Quánzhōu was one of the most cosmopolitan cities in the world during the Sòng and Yuán dynasties, a city that was temporarily home to many experienced foreign (mostly Arab) navigators who would exchange information and their knowledge of ocean navigation among themselves and with their Chinese hosts.

We learn from a hand-written copy of the *Notebook on Sea Bottom Currents*, found in Quánzhōu, that in 1382 the emperor ordered the Islamic observatory official Hay-dar (Hǎi dá er; 海答兒) and Mashā'ikh (Mǎ shā yì hēi; 馬沙亦黑), a master of Islam, to collect the best astronomy books from among several hundred volumes of the *Books from the Western Regions* at the Yuán court. In 1383 their selection was published in Chinese as the *Works of Astronomy* (*Astrology*). The volume included a discussion of the Islamic concepts of longitude and latitude. The *Works of Astronomy* was originally composed by the mathematician Abu Hassan Koshiya (971–1029), who played an important role in the development of spherical trigonometry.

In 1403 Zhèng Hé represented the Yǒnglè Emperor, Zhū Dì (朱棣; 1360–1424), on a mission to recruit foreign (Arab) navigators for transoceanic voyages and collect information about sea currents, islands, mountains, straits and the positions of stars so as to revise their navigation charts. The Yuán Emperor's focus on enlisting Arab astronomers is obvious from the passage below, which was taken from the *Treatise on Calendrical Astronomy from Western Areas* by the Imperial official Xú Yǒuzhēn (徐有貞; 1407–1472):

> The Muslim system of calendrical astronomy is said to have been authored by an extraordinarily talented man named Mahamute [Mohammed] in the *Aerbi* [Arab] year, in a Western place called *Make* [Mecca] ... Calendar makers believed it to be the most accurate system for the calculation of the longitudes, latitudes as well as ominous motions [of the Seven Luminaries, that is, the seven moving astronomical objects in the sky visible to the naked eye: the Moon, Mercury, Venus, the Sun, Mars, Jupiter and Saturn], but it did not spread eastwards before the Yuán era.

When the first Emperor [of the present dynasty] inaugurated the composition of the *Dàtŏng lì* [the 'great universal system of calculating astronomy'], he summoned to his court [several] Westerners, experts in calendar making, and ordered the Bureau of Astronomy to adopt the Muslim astronomical techniques as a supplement [to the *Dàtŏng lì*]. It has remained in use since then. My friend Liú Zhòngfú [劉仲孚; *viz*. Liú Xìn 劉信] has excellent knowledge of stars and astronomical tables. He has comprehensively studied all sorts of [Chinese astronomical] methods and is also proficient in Western [Islamic] ones. He noted that the Muslim system of calendrical astronomy is somewhat inconsistent and devoid of uniform rules, and would therefore become more and more confusing with the passage of time. Consequently, he translated the Muslim text precisely, prescribed rules for its usage and pre-calculated the essential quick tables. The resulting procedures are brief, simple and clear. They form an orderly book devoted to the [Muslim] school of astronomy that is to be used indefinitely and to remain essential to students of calendrical astronomy. (Xú, 1439: 21a–23a; translation: Shi, 2003: 32).

## 7 ADMIRAL ZHÈNG HÉ

### 7.1 Guided by astronomy

Other transoceanic voyages undertaken in the early fifteenth century may not have had the need to rely on their navigators' skills to determine longitude at sea. For instance, the fourth expedition (1413–1415: see Figure 7, left) of the eunuch Admiral and Chief Envoy Zhèng Hé set off from Sumatra, in present-day Indonesia, sailing 10 days in full wind to reach the Maldives archipelago. From there, they sailed across the Indian Ocean in full wind for 15 days, reaching Mogadishu in East Africa, some 3700 miles from their port of origin (Wang, 2004a; Morton and Lewis, 2005: 128, translation):

> We have traversed more than 100,000 *lĭ* [50,000 km] of immense water spaces and have beheld in the ocean huge waves like mountains rising in the sky, and we have set eyes on barbarian regions far away hidden in a blue transparency of light vapours, while our sails, loftily unfurled like clouds day and night, continued their course [as rapidly] as a star, traversing those savage waves as if we were treading a public thoroughfare… (From a tablet erected by Zhèng Hé in Chănglè, Fújiàn, in 1432; see also Duyvendak, 1939: 345).

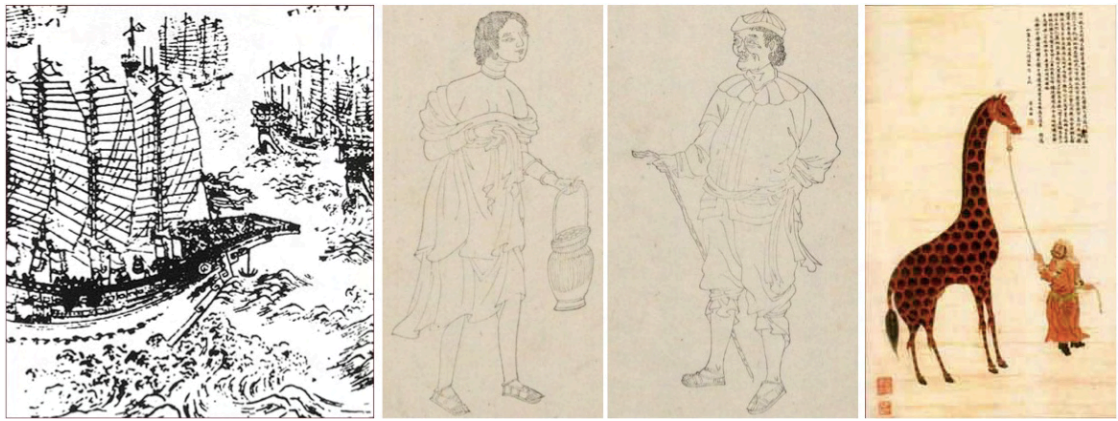

**Figure 7.** *(left)* Zhèng Hé's fleet (Chinese woodblock print, via Wikimedia Commons; public domain). *(middle panels)* African tribute-bearing ambassadors to the Qīng dynasty (1761) (public domain). *(right)* Tribute giraffe from Bengal (National Museum of China; public domain).

This voyage appears to have followed well-established Arab and Chinese trade routes rather than breaking new navigational ground. Dreyer (2007: 35–36) clearly explains the navigational premise of Zhèng's voyages (my emphasis):

> All of the voyages were within waters dominated by the seasonal pattern of monsoon winds. From December to March, high pressure over Central Asia produces cold, dry winds that blow from the north and the northeast ... This weather pattern reverses itself in April, when winds laden with vapor arise in the subequatorial waters of the Indian Ocean. Blowing northward in a generally circular pattern, they approach India and Southeast Asia from the south or southwest ... **Predictable monsoon winds also aided navigation.** With **latitude calculated fairly easily from stellar observations**, predictable monsoon winds gave navigators the confidence to sail straight from the southern Arabian peninsula to the west coast of India, or from Ceylon and southern India to Sumatra and the Malay Peninsula, following a particular latitude line. Well before the rise of Islam, Arabs from the southern part of the peninsula had discovered this secret, and they were joined on the ocean by Tamils from southern India and Indonesians from Sumatra and other islands.

In his discussion of Zhèng's seventh voyage (19 January 1431–22 July 1433), Dreyer (2007: 161) explicitly refers to the practice of 'running down the latitude', which was also adopted by early European seafarers (e.g., de Grijs, 2017: 2-47), including Columbus and Vespucci (my emphasis):

> The fleet spent three full days at Zhan City and then set sail on 17 June [1433], the first day of the sixth lunar month. The *Xià Xīyáng*[10] [*Down to the Western Ocean*] records several sightings on the next leg of the voyage, incidentally providing confirmation that the navigators of Zhèng Hé's fleet were happy to sail by landmarks when they could find them, **rather than by dead reckoning** [sailing by knowing one's latitude and compass course while accounting for the effects of the wind and currents].

Nevertheless, Zhèng routinely travelled to Quánzhōu, as well as to the southern Chinese provinces of Fújiàn, Guǎngdōng and Zhèjiāng, prior to each of his voyages in search of talented foreign navigators residing in China, who would be rewarded handsomely upon their return from successful missions. It is conceivable that Zhèng's navigators may have used sightings of the rising or setting Sun and Moon to help determine how far the ship had sailed in an easterly or westerly direction (which is equivalent to determining longitude based on an approach similar to the lunar distance method), while they used stellar elevations to determine both their longitudes and latitudes (Gong, 2005).

On 10 October 1432, Zhèng's fleet set sail once again, this time from Pulau Rondo on Sumatra to Sri Lanka. The fleet continued to Kuli (Calicut) in India, where they arrived on 10 December 1432. Their next leg commenced three days later and involved a 35-day voyage, until 16 January 1433, to Bandar-e Abbas in the Strait of Hormuz. They soon proceeded with their next transoceanic leg, crossing the Arabian Sea from Dingde Baxi (present-day Dandi Bandar) to Jabal Khamis in present-day Oman.

The expedition used the declination of *Zhīnǔ* (the 'weaving girl,' corresponding to the star Vega) and of *Nánmén shuāngxīng* (two stars in the constellation of Sagittarius) for geographic location determination. Travelling westwards, they relied on observations of Pollux and Procyon to determine their longitudes and latitudes (Liu, 1989; Wang, 2004b: 395–397; Liu et al., 2005). Specifically, Zhèng's navigators based their transoceanic voyages on 10 stars, which enabled them to determine their approximate longitudes and latitudes at sea. For latitude determination and navigation, they relied on stars close(st) to the North and South Poles, respectively Polaris and the Southern Cross combined with Canopus, in turn cross-referenced to Polaris (for details, see Liu, 1989; Liu et al., 2005; Ding et al., 2007), as well as on stars in Sagittarius. For longitude determination, their guide stars included Pollux in the northwest, Procyon in the southwest and Vega and bright stars in Taurus in the East (Wang, 2004b: 396).

Ding et al. (2007) pointed out that Zhèng and his navigators used star boards rather than sextants (see also Jin, 1996), although both types of tools serve the same purpose. Zhèng had access to 12 different star boards, whose use would depend on the altitudes of the stars of interest (e.g., Jin, 1996):

> The size of the largest board [was] around 24 cm × 24 cm (12 fingers …, using the size of [the] human thumb) and the smallest, 2 cm × 2 cm (1 finger), with [successive differences of] 2 cm [each] (Zhao, 2005). Using the attached thread, around 60 cm in length, the observer was able to fix the distance from the eye to the board, thus fix the angle of observation of the board, about 1.9° per finger. When observing celestial bodies, the observer aligned the star [with] the [top] of the board and the horizon [with] the [bottom], and thus the height of the star was determined. (Ding et al., 2007: Section 3.2).

The resulting positional uncertainty was as small as 4.5 nautical miles (Zhao, 2005; although it is unclear to which latitude range this uncertainty applies), an unprecedented accuracy for the times. Since the *Dàtǒng lì*—which was based on Guō Shǒujìng's pioneering system of calendrical astronomy—had been formally adopted by the Míng Bureau of Astronomy in 1384, it is conceivable that Zhèng and his navigators were familiar with longitude determination at sea on the basis of solar and lunar eclipses as well.

**7.2 Zhèng Hé's navigation tools**

In his capacity of senior administration official, between 1405 and 1433 Zhèng led seven expeditions to southeast, South and West Asia, and to East Africa. They were aimed at establishing a Chinese presence, enforce Imperial control over trade in the Indian Ocean and extend the Yǒnglè Emperor's tributary territories (see Figure 7, right and middle panels). Zhèng's sailing charts are thought to have consisted of four maps, one each centred on Sri Lanka, South India, the Maldives and a fourth map showing some 400 km of the East African coast, covering the coastal waters to a southern latitude of six degrees.

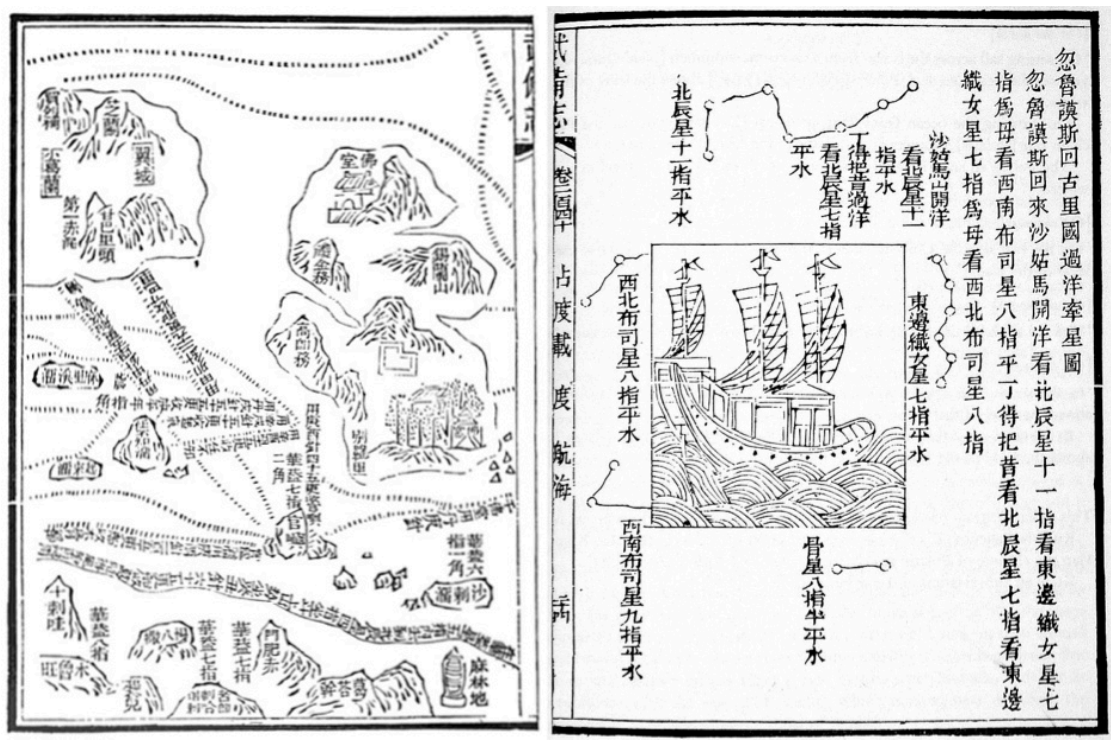

**Figure 8.** *(left) Máo Kūn Map* showing Sri Lanka (main island on the right) and the east coast of Africa (bottom). India is located on the left and Malindi is indicated in the far right-hand corner. (Mills, 1970: Appendix 2.) *(right)* Star charts (ca. 1628) based on the *Máo Kūn Map*. The chart specifies the guiding stars *en route* from Hormuz to Calicut. (Mills, 1970: Chapter 240, folio 24r.)

Zhèng's maps contained sailing directions for more than 40 oceanic trade routes (Ding et al., 2007; Sun, 2009). Note that the *Máo Kūn Map* (Figure 8), which is usually considered a good representation of Zhèng's map, was in fact compiled approximately two centuries after Zhèng's voyages, and so the *Máo Kūn Map* may not accurately represent the cartography available to the Admiral. Modelled on the *Great Chart of the Yangtze River*, the *Máo Kūn Map* was originally a scroll collection that had to be read from right to left. The scroll, the first international chart in existence anywhere in the world, was titled *Chart of the course taken by the treasure ship from its start at Lóngjǐng to the destinations of foreign lands* (Ding et al., 2007). It was subsequently converted into a 24-page book, including two pages containing four navigational star charts, when it became part of the *Treatise on Armament Technology* (e.g., Sun, 2009).

Zhèng's sailing charts were designed—likely with the help of Arabic-speaking pilots with detailed knowledge of the African coast—for the specific purpose that they would be used to sail along certain routes. As such, their positions vary in orientation to align with the ocean currents and winds, as required of a sailing chart, specifically from the user's perspective (e.g., Sun, 2009). The aim was clearly to provide positional information in the shortest time possible, as also evidenced by the orientation of the geographic features: they are seen from the user's orientation, and so they do not feature coordinate graticules nor scales, and neither are they oriented according to fixed cardinal directions (e.g., Sun, 2009). The margins of the coastal charts contained a selection of realistic drawings of coastal features—including peaks, islands, rocks, temples, bridges and buildings.[11]

Sailing instructions were offered using a 24-point compass system with a Chinese symbol for each point, combined with a sailing time or distance. Local currents, winds and depth soundings were taken into account. For instance, off the coast of China's Zhèjiāng and Fújiàn provinces, we read

> Ships sail across the Xiào Shùn Ocean and arrive at Jiú Shān [Shān = Mountain], where the depth of the water is 9 *tuǒ* [one fathom; two outstretched arms equal 1 *tuǒ* or 1.66 m]. In the south-west on the opposite side from Jiú Shān, there are sunken rocks where great waves break over them ... In the north-west there are sunken rocks where great waves break over them, be careful when sailing ... Dōng Luò Shān, pass through the channel between the islets ... Xiá Mountain, pass outside. (Sun, 2009: 338).

This level of detail appears to have been common, given that Zhèng's other navigational chart collections, e.g., for the waters surrounding the Maldives archipelago, Gǒng Zhēn's (鞏珍; fl. 1430–1434) *Accounts of the Western Countries* (1434) warns,

> … apart from the eight main islands where trade could be carried on ... there are 3000-odd places where there are eddies of water. Keep clear of them carefully lest ships be wrecked. (Sun, 2009: 338).

Despite the absence of a latitude and longitude grid, a physical scale or a north-pointing bearing, "… the errors [were] usually not more than 5 degrees, which is thought to be wonderful for the helmsmen of 1425." (Needham, 1971, cited by Sun, 2009: 337).

The 24-point compass system resembles the system used for navigation in the western Pacific, which was used by the ancient Polynesian navigators (e.g., de Grijs, 2017: Chapter 2). In a Chinese cultural context, the compass points are based on the 12 directions of the *Earthly Branches*, a system for timekeeping. For navigation purposes, 12 points were insufficient, thus prompting the introduction of 24- or even 48-point compasses by inserting intermediate directions. Zhèng's fleet navigated on the basis of both so-called 'wet' compasses—composed of a needle floating in water contained in a circular box with the compass points carved on wooden rim—and magnetic compasses (Ding et al., 2007).

As early as the Yuán and early Míng Dynasties, the Chinese had already invented an approach to timekeeping that was reasonably accurate and worked on land as well as at sea. They were indeed far ahead of European developments, which would not bear fruit until the seventeenth and eighteenth centuries (e.g., de Grijs, 2017: Chapters 3–6). A particularly promising Chinese timing device was the everbright lantern, which operated similarly to the common candle clocks in use at the time. Xiè Jié (謝杰), a contemporary Míng author, elaborated on what was known as the compass navigation method of the Zhāngzhōu people:

> The compass cabin burns everbright lanterns day and night, five *gèng* each night and five *gèng* each day. So a ship sailing for 12 Chinese hours [24 hours] burns the equivalent of a total of 10 *gèng*. (Han, 2002: 248; translation: Wang, 2011: 2).

The accuracy of this timekeeper was remarkable for the times. Over the course of 24 hours, or 10 *gèng*, a lantern would consume one catty of oil (Han, 2002: 248–249), equivalent to approximately 500–600 grams, so that its burn rate could be determined to better than 0.1 catty by reading off the amount of oil consumed on a graduated glass reservoir. Chinese navigators could thus keep track of the local time in their homeport while out at sea. Comparing the ship's local times of sunrise or sunset with the homeport's time based on the lantern, they could obtain a rough estimate of their longitude difference.

Incense coils with known, carefully calibrated combustion timescales had been used for the same purpose since the sixth century and became commonplace in the Sòng dynasty (Richards, 1998: 52). Both timing devices were used for periods of up to weeks or even months. To obtain a more accurate measurement of the time (and longitude) difference with respect to the homeport, water clocks were employed to time the difference between a homeport's benchmark time (e.g., the time of sunrise) and that at the ship's location. This was done most accurately by weighing the amount of water consumed between both calibration moments (Han, 2002: 248–249).

After the death of the Yǒnglè Emperor in 1424, Zhèng's expeditions were discontinued and a period of decline in cartographic developments commenced. China began to look inwards, adopting a policy of isolationism that lasted a few hundred years. The new Hóngxī Emperor (1378–1425) ordered that Zhèng's fleet be burnt, along with all records, thus ending the 'Age of the Sea.' In addition, the Hóngxī Emperor expressly prohibited overseas travel. Anyone who disobeyed the order was killed.

**8 JESUIT INNOVATIONS**

Traditional Chinese cartographic skills became more advanced under the influence of new ideas introduced by the European Jesuit missionaries in the late Míng dynasty, from the late sixteenth century onwards. However, major new initiatives were not seen until well into the Qīng dynasty (1644–1912), when the Kāngxī Emperor (1654–1722) realised that Chinese maps were not sufficiently accurate for navigation and territorial purposes. Therefore, in 1708 he sponsored a geodesy and mapping programme using astronomical observations and triangulation measurements (Hostetler, 2013). The final product, the first on-the-spot survey map, the *Kāngxī Provincial Atlas of China* (1721/2), took well over a decade to complete.

**8.1 Matteo Ricci**

Matteo Ricci (1552–1610) was the first European Jesuit to take residence in mainland China, in 1583. In the context of the longitude story, Ricci's most important contributions to Chinese developments include his unwavering efforts to produce accurate world maps, his *mappa mundi*. Following the first edition of his *Complete Map of the Earth's Mountains and Seas*,[12] published in 1584, Ricci produced a total of eight editions of this map between 1584 and 1608.

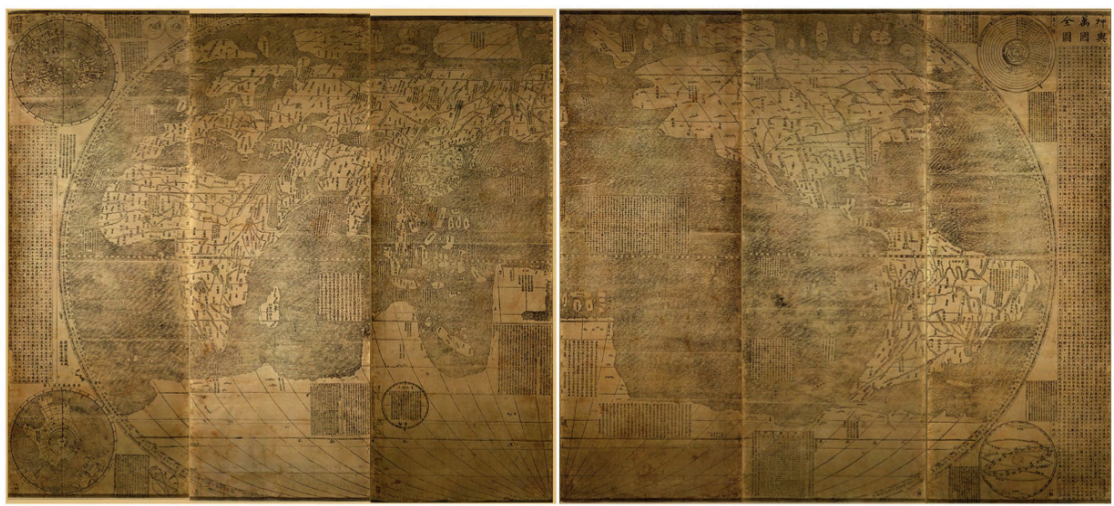

**Figure 9.** Composite of Matteo Ricci's *Map of the Myriad Countries of the World* (1602). (Courtesy: Library of Congress, Geography and Map Division, digital IDs g3200.ex000006Za, b.)

The map's third edition, the *Complete Map of the Myriad Countries on Earth* (see Figure 9), was produced in 1602 either by or with the help of the powerful Míng official Lǐ Zhīzǎo (李之藻; 1565–1630), who had been Ricci's student since 1601 (e.g., Elman, 2007; Hart, 2013: 253). Ricci's maps combined quantitative cartography by means of a carefully drawn longitude and latitude grid with a prominent narrative component. In the map's right-hand margin, Ricci introduced the prevailing Catholic interpretation of the cosmos as well as extant European geographical knowledge of astronomy, cartography, world geography and Aristotelian scholastic natural philosophy (e.g., Zhang, 2015). In addition to these descriptions, which included overviews of the five continents, the world's oceans and the different climate zones, Ricci also explained the causes of solar and lunar eclipses, the relative sizes of the Earth, the Sun, the planets and the fixed stars as well as their respective distances, and the translucency and solidity of the crystalline celestial spheres.

Ricci's description of the Earth as a globe, a 'terrestrial sphere', located at the centre of the cosmos prompted a revision of Chinese ideas of the world at large. He introduced the concepts of the Equator, the North and South Poles, and the identification of locations by means of their latitudes and longitudes. He even attempted to provide an estimate of the size of the Earth. From personal observation, he initially suggested that one degree of latitude corresponded to 200 *lǐ*; in turn, that measure of length implied that the Earth's circumference was 72,000 *lǐ* or about 36,000 km (Zhang, 2015: note 45). Around 1600, Ricci revised his estimate to 250 *lǐ* per degree latitude, or a circumference of 90,000 *lǐ* (45,000 km). In 1702, in the late Kāngxī reign, Míng literati reversed the estimate once again, to 200 *lǐ* for one degree of latitude.[13] Based on our present knowledge of the Earth's size, its circumference is 40,075.017 km (80,150.034 *lǐ*), corresponding to 222.63 *lǐ* per degree latitude.

Most important from a practical perspective, Ricci and his successors introduced Western knowledge of mapping to Chinese scholars, including surveying and methods using degrees of latitude and longitude. Nevertheless, despite Ricci's introduction of mathematical geography and his influence at the Míng court, the traditional Chinese grid system continued to dominate Chinese cartography throughout the late Míng and Qīng periods (Elman, 2007). However, some aspects of Ricci's teachings gradually found their way into local map-making. For instance, the scholar Cáo Jūnyì (曹君義) produced a 'Complete Map' in 1644 that mixed Chinese classical approaches to map-making with reasonably accurate locations of Europe, Africa, India and Central Asia based on the use of longitudes to determine the relevant distances from Nánjīng (see Smith, 1998: 71–83). However, as late as in 1832, Lǐ Zhàoluò's (李兆洛; 1769–1841) atlas of the Qīng Empire showed both grid and latitude and longitude lines on the same map (Elman, 2007), thus highlighting the reluctance of home-grown cartographers to fully relinquish the traditional system.

**8.2 Seventeenth-century developments**

Later Jesuit missionaries to China built on Ricci's early work. Specifically, they reinforced the classical idea of a rotating, spherical Earth, which in turn implied that longitude differences were equivalent to time differences. The Jesuits' main focus was on contributing to astronomical measurements of solar and lunar eclipses, the ephemerides of Jupiter's satellites, comets and transits of Mercury across the solar disk. Those observations were all employed to determine accurate longitudes and latitudes across China and to obtain firm constraints on the shape of the Earth.

Among the Chinese scholars in the Jesuits' orbit, the scholar and Imperial official Xióng Míngyù (熊明遇; 1579–1649) was the most senior member. He befriended a number of Jesuit missionaries around 1613–1615, including Sabbathin (Sabatino) de Ursis (1575–1620). Xióng offered support to the notion that the Earth was spherical in his preface to de Ursis' treatise *Explanation of the Gnomon* (1614)—as well as in his own *Manuscript on the Investigation* [*of the Principle of Things*] *to Attain* [*Knowledge*] (1648)—mostly because he was convinced that the Europeans had offered the requisite proof by having circumnavigated the globe (Zhang, 2015: Chapter 4). In a section of the latter treatise titled *There Are No Vacuums or Obstacles on the Round Earth*, Xióng explained,

> In fact, navigators can sail around the Earth as if making a circle … using longitudes running North to South and latitudes [such as the] Equator. They measure the position of the Sun during the day and of the stars at night and are able to determine the high or low [of their location] and the distance they have travelled. Thus the Earth can be navigated. (Xióng, 1648: 151b–152a; translation: Zhang, 2015: 54).

Thus far, Chinese navigators had only been able to determine their latitude at sea (and not their longitude), given the absence of a cultural notion allowing for the Earth's sphericity. Xióng's new insights represent the first steps by a Chinese scholar to embrace that latter concept.

The increasing availability of accurate lunar eclipse timing predictions [14] and of ephemeris tables allowing the calculation of immersions and emersions of Jupiter's Galilean satellites, such as those published in 1683 by Giovanni Domenico Cassini (1625–1712), Director of the Paris Observatory, greatly enhanced the Jesuits' productivity. Observations of the immersions and emersions of Jupiter's satellites were the preferred approach (Goüye, 1688: 232) to determining longitudes given their frequent occurrences. However, owing to the prevailing timing uncertainties of about 2 minutes in Cassini's ephemeris tables, the inherent uncertainties in contemporary longitude determinations were at least half a degree.

Gislén (2017) analysed the quality of Chinese ephemeris observations obtained in 1689–1690 and the longitudes thus derived. He concluded that the observations were of high and consistent quality, and that the resulting longitudes of Hoai-ngan (Huái'ān), Si-ngan-fu (Xī'ān), Canton (Guǎngzhōu), Chang-hai (Shànghǎi), Nankin (Nánjīng) and Peking (Běijīng) were in good agreement with modern values. However, one should keep in mind that the Jesuits had obtained their carefully calibrated geographic location measurements on solid ground; doing the same on board pitching and rolling ships was a different matter altogether.

The most important astronomical and geographical contributions were made—both *en route* to China and in residence—by Jean de Fontaney (1643–1710), Antoine Thomas (1644–1709), Guy Tachard (1651–1712), François Noël (1651–1729), Jean-François Gerbillo (1654–1707), Louis Le Comte (1655–1728), Joachim Bouvet (1656–1730), Claude de Visdelou (1656–1737), Ignaz Kögler (1680–1746), Jean-Baptiste Jacques (1688–1728), André Perreira (1689–1743), Antoine Gaubil (1689–1759) and August von Hallerstein (1703–1774) (e.g., Udias, 1994, 2003; Uhalley and Wu, 2015: 142; Gislén, 2017).

**8.3 The Kāngxī Jesuit Atlas**

In 1708, the 47th year of his reign, the Kāngxī Emperor commissioned a nationwide geodetic survey. He also commissioned the production of a comprehensive map of his empire, which would be based on longitudes and latitudes rather than on Péi Xiù's 'six principles of cartography', as had been advised by the classicist scholar Hú Wèi (胡渭; 1633–1714):

> Alas! Since the collapse of the Jìn dynasty cartography has become an extinguished learning. How pitiable! I wish there were someone willing to take up the mission to restore ancient learning and memorialise the throne to obtain a decree making all local magistrates survey the roads in their jurisdictions by land and water, take note of the high and low, the precise spatial alignment, and the curving and straight, and submit all these data to the highest official in charge of revenue. This official then would hire mathematicians throughout the country to calculate the correct horizontal distances between these places and, noting their correct positions relative to each other, would then put all the data into a book. …The maps made on the basis of such data would show all the places correctly; there would be no attaining accuracy in one corner and losing it in another. [If that happens,] then Péi Xiù's extinguished learning would be revived again. Wouldn't that be an achievement worthy of celebration for ten thousand generations? (Hú, 1701: 123–124; translation: Zhang, 2015: 249).

The Kāngxī Emperor's ultimate aim was to unify the entire country and strengthen the border regions against foreign aggressors by producing a comprehensive, accurate and detailed national map. He ordered the Jesuits to collect local maps, purchase instruments and recommend specialists to engage with. Meanwhile, he established a national geographic survey and mapping framework. Over the course of the next ten years, survey teams experienced considerable hardships while climbing mountains and crossing streams to survey the major locations of interest in each province; they engaged local officials and also referred to local historical documents. Latitudes and longitudes were determined based on astronomical surveys, adopting the longitude of the Qīng capital, Běijīng, as the prime meridian.

Kāngxī's Jesuit Atlas was eventually completed in 1718. The central map, the *Map of a Complete View of the Imperial Territory*, drawn at a scale of 1:1,400,000 and using the so-called 'trapezoid' projection, was

> … very detailed, including strategic junctures and forts, dikes, coastal defences, village fortifications and garrisons, postal stations and ferry towns, strategic defiles, environment, defenses and transportation, no matter how remote and desolate. (*Veritable Records of the Kāngxī Reign*, chapter 283; translation: Y. Chen, 1996: 321).

The renowned sinologist Joseph Needham (1985: 235) has praised the map as "…not only the best of all the maps which had ever been made in Asia but better and more accurate than any European map of its time." Based on the Kāngxī Jesuit Atlas, combined with ancient Chinese records, the French astronomer Alexandre-Gui Pingré (1711–1796) calculated Běijīng's longitude and latitude; his results, published in 1764, became the standard geographic reference.

However, the excellence of the Kāngxī Jesuit Atlas was lost on the majority of scholars. The Qīng government considered the map a treasured state secret and consequently restricted access, keeping it hidden in the Imperial Palace. Outsiders were not provided with any details, and even reference materials in Chinese were scarce (Y. Chen, 1996). Although Chinese achievements in surveying and mapping had reached the highest and most advanced level for the times, Chinese cartography did not continue on this promising, upward trajectory.

With the withdrawal of Imperial support and a renewed period of inward focus, no new cartographers were trained in the field's state of the art and progress ceased. In addition, production of planimetric maps like the *Golden Mirror of Channelling Water*, which were produced around the same time as the Kāngxī Jesuit Atlas, to some extent using similar principles as the Jesuits' maps (e.g., Yee, 1994), was no longer supported. In fact, the modern cartographic methods introduced by the Jesuits therefore only had a limited impact on Chinese cartography, and so most of Chinese society continued to believe that the Earth was flat (Desnoyers, 2004; Zhu, 2001). It took until the late Qīng era, during the Guāngxù period (1871–1908), before cartographic developments received renewed impetus and general societal attitudes started to change (Zhu, 2001). Maps were finally drawn based on longitude and latitude surveys, using modern map projections.

## 9 FINAL THOUGHTS

Contrary to the dominant sea-faring nations in western Europe, historical cartographic developments in China were less focussed on finding one's position at sea than on accurate representation of local areas. While European nations pursued oceanic exploration driven by powerful, persistent economic motivations, such conditions were almost entirely absent in China. Until the arrival of the first Jesuit missionaries in the late sixteenth century, the Chinese worldview was narrow and the notion of the Earth as a sphere had not yet found common ground. With the exception of Zhèng Hé's fabled voyages aimed at securing loyalty from and submission of foreign leaders, the Chinese did not undertake any major exploratory expeditions.

Nevertheless, Zhèng Hé's expeditions demonstrated that Míng-era China did have a powerful capacity for ocean navigation. To some extent influenced by innovations introduced by Arab navigators (e.g., Zhang, 2002), the Chinese developed their own independent means of longitude determination, based on both astronomical observations and cleverly designed timekeeping devices. Zhèng Hé's maps, while different in orientation and design from their European counterparts, reflect the high level of navigational expertise and maritime technology already achieved and available at the time.

Chinese maps were not uniformly 'Westernised' until at least the end of the nineteenth century. However, from the Bureau of Institutional Studies (Yee, 1994) we learn that combining maps from different provincial governments into a large-scale national atlas was complicated by the wide diversity of map-making methods in use, despite the official adoption of nationwide cartographic standards:

> When [new maps] were compared with old accounts, there were still discrepancies and similarities. These were all checked by measuring distances and by widely searching through records in an effort to resolve ambiguities and reach a single conclusion. (*Qīndìng dà qīng huì diǎn*, 1899: 2, 1022; translation: Yee, 1994: 126).

Text-based verification and narrative descriptions thus remain integral components of Chinese cartography, even today.

# 10 NOTES

[1] References to 'China' and 'Chinese' are used as shorthand to identify the general area encompassed by the modern People's Republic of China. However, I do not intend to strictly apply present-day national borders to my narrative.

[2] Where Chinese names are used, I have adopted the common convention of ordering these names as 'surname, first name'.

[3] Figure 1 is a portrait gallery of the main characters driving the developments described in this paper, if and when their images were available in the public domain. I encourage the reader to refer to this compilation figure whenever a new character is introduced in a more than cursory manner.

[4] Where the source of a translation has not been included explicitly, the English text was taken from the reference cited.

[5] 'Zhāng Héng' was formerly transcribed as 'Chāng Héng'. He is also known as 'Píngzhì'.

[6] Yabuuti and van Dalen (1997: footnote 2) point out that in the *Calendar Annals* of the *History of Yuán* Ghengis Khan's western expedition is misidentified as the "… western expedition of Ögödei …", Ghengis' son and successor.

[7] Note that the modern *pīnyīn* (English transliteration) rendition of this Chinese name is Liú Chāo.

[8] Owing to the combination of the Earth's rotation and its orbital motion around the Sun, the apparent time of the passage of any given star through the local meridian changes by about 4 minutes from day to day. For each day in the full 1461-day cycle, one could then produce ephemeris tables for stars passing through the prime meridian at a given time. At sea, a navigator can determine the time of the passage of a specific star through the local meridian. Compared with the star's tabulated time of passage through the prime meridian, a time difference and, hence, a longitude difference follows.

[9] Note that the Chinese adopted a system composed of 365¼ degrees of longitude versus the 360 degrees in use in Europe. In addition, Chinese latitudes were determined with respect to Polaris, while Europeans used the Equator as their reference.

[10] The *Sānbǎo tàijiàn **xià xīyáng** jì tōngsú yǎnyì* (*Romance of the Sānbǎo Eunuch's Voyage to the Western Ocean*; 1597) is a novel in one hundred chapters covering Zhèng Hé's seven voyages of 1405–1433. It is usually attributed to Luó Màodēng (羅懋登; fl. 1596).

[11] A similar approach to cartographic orientation was adopted for some early Western nautical chart collections, including those of the famous Dutch cartographer Lucas Janszoon Waghenear (ca. 1534–1606)—of *Spieghel der zeevaerdt* (*Mariner's Mirror*; 1584) and *Thresoor der zeevaert* (*Treasure of Navigation*; 1592) fame—and his compatriot, the portrait painter Cornelis Anthoniszoon (ca. 1505–1553). Their charts "… were drawn with the general direction of the coast stretching from left to right [i.e., in the opposite orientation compared to that adopted by the Chinese] across the sheet without reference to geographical North." (Cotter, 1980: 425).

[12] The genre of *Complete Maps of All under Heaven* was initiated by Huáng Zōngxī (黃宗羲; 1610–1695).

[13] The Kāngxī Emperor personally decreed that the ruler produced by the Ministry of Works was to be adopted as the national standard of length. The resulting scale implied that five *chǐ* (1 *chǐ* = 1/3 m) was equal to one *bù*, 360 *bù* = 1 *lǐ* and one degree of latitude = 200 *lǐ* (Zhang, 2015: Chapter 4).

[14] Around the turn of the eighteenth century, Isaac Newton (1642–1726/7) had developed his *Theory of the Moon's Motion* (1702). The underlying ideas eventually found their way to China, but although Newton's *Theory* was used in Europe to determine longitudes, in China it was only applied to predict lunar eclipse timings (Kollerstrom, 2000).

**11 ACKNOWLEDGEMENTS**

I am grateful to Nà Jié and Chén Xiàodiàn for their generous assistance in helping me check my renditions of traditional Chinese characters and *pīnyīn* phrases throughout this manuscript. I also thank the librarians of Western Sydney University's Joyce Wylie Library (Bankstown) for providing me with access to essential physical reference materials despite stringent COVID-19 restrictions on external visitors.